\documentstyle[11pt,epsfig]{article}

\newcommand{\bee}{\begin{equation}}
\newcommand{\ene}{\end{equation}}
\newcommand{\bea}{\begin{array}}
\newcommand{\ena}{\end{array}}
\newcommand{\beqa}{\begin{eqnarray}}
\newcommand{\enqa}{\end{eqnarray}}
\newcommand{\bean}{\begin{eqnarray*}}
\newcommand{\eean}{\end{eqnarray*}}

\def\ru1{\rule[-0.4truecm]{0mm}{1truecm}}

\def\ol{\overline}
\def\bB{\overline{B}}

\def\up#1{\leavevmode \raise.16\hbox{#1}}

\def\sqr#1#2{{\vcenter{\vbox{\hrule height.#2pt
	\hbox{\vrule width.#2pt height#1pt \kern#1pt
	  \vrule width.#2pt}
	\hrule height.#2pt}}}}

\newcounter{appendix}

\def\thebibliography#1{{\bf REFERENCES\markboth
 {REFERENCES}{REFERENCES}}\list
 {[\arabic{enumi}]}{\settowidth\labelwidth{[#1]}\leftmargin\labelwidth
 \advance\leftmargin\labelsep
 \usecounter{enumi}}
 \def\newblock{\hskip .11em plus .33em minus -.07em}
 \sloppy
 \sfcode`\.=1000\relax}

\begin{document}

\title{\hfill $\mbox{\small{\begin{tabular}{r}
${\textstyle Bari-TH/98-298}$\\
${\textstyle DSF-T-10/98}$\\
${\rm\textstyle April~~1998}$
\end{tabular}
}}$ \\[1truecm]
PHENOMENOLOGICAL BOUNDS ON B TO LIGHT SEMILEPTONIC FORM FACTORS } 

\author{
D. FALCONE$^{\ast}$, 
P. SANTORELLI$^{\diamond}$, 
N. TANCREDI$^{\ast}$, 
F. TRAMONTANO$^{\ast}$ 
} 
\date{$~$}

\maketitle

\thispagestyle{empty}

\begin{center}
\begin{tabular}{l}
$^{\ast}$ 
Dipartimento di Scienze Fisiche, Universit\`a di Napoli, 
Mostra d'Oltremare, Pad.19,\\
~~I-80125, Napoli, Italy; INFN, Sezione di Napoli, Napoli, Italy. \\
$^{\diamond}$ Dipartimento di Fisica, Universit\`a di Bari, 
Via G. Amendola, 173, I-70126 Bari, Italy; \\
~~INFN, Sezione di Bari, Bari, Italy.
\end{tabular}
\end{center}

\begin{abstract}
The form factors for the weak currents between B and light mesons are
studied by relating them to the corresponding D form factors at
$q^2_{max}$ according to HQET, by evaluating them at $q^2=0$ by QCD sum
rules, and by assuming a polar $q^2$ dependence. The results found are
consistent with the information obtained from exclusive non-leptonic 
two-body decays and, with the only exception of $A_1$, with lattice
calculations. 
\end{abstract}

\bigskip


\newpage


The properties of hadrons containing a single heavy quark $Q$ are
characterized, in the limit $m_Q \to \infty$, by $SU(2n_Q)$ spin-flavour
symmetry of the strong interactions \cite{iwgeorgi}. This symmetry
allows us to understand the spectroscopy and the decays of heavy hadrons. In
particular for the $\ol B \to D^{(*)}~l^-~\ol{\nu}$ transitions all the
form factors are proportional to the Isgur-Wise function \cite{IWf},
$\xi(w)$ where $w=v_B \cdot v_{D^{(*)}}$, which is equal to 1 at
zero-recoil point ($w=1$). Thus the Heavy Quark Effective Theory (HQET)
has been successful in obtaining a rather precise determination of the
modulus of the CKM matrix element $|V_{cb}|$ from the study of the
semileptonic decays \cite{neubertsem}. This task was obtained also
considering symmetry breaking and perturbative corrections
\cite{QCDcorr}. 

Another consequence of the Heavy Quark Symmetry (HQS) is the possibility to
relate the semileptonic heavy-to-light form factors; indeed the
form factors $B \to L$  (L indicates a light meson) are related to the
ones of $D \to L$ at the same velocity ($v_B=v_D$) \cite{IW2}. The
symmetry, however, does not help us to fix their normalization as
it is the case for the heavy-to-heavy form factors. However, from the
phenomenological point of view, the knowledge of the heavy-to-light form
factors is fundamental to determine, for example, the magnitude of
the $|V_{ub}|$ CKM matrix element \cite{Ball1}. The rare B decays
represent an important channel for testing the standard model
predictions and also in this case the knowledge of the heavy-to-light
form factors is the fundamental ingredient of the investigation
\cite{Burdman}. 

Here we study the $q^2$ dependence of the heavy-to-light form factors. 

The heavy flavour symmetry relates charmed semileptonic form factors to those
of the beauty particles \cite{IW2}. Therefore, in principle, we can
evaluate the $B \to L$ form factors from the experimental knowledge of
the form factors appearing in the semileptonic decays of the D mesons
\cite{Exxx}. This approach was used in \cite{io,Tanimoto,BLMS} to
estimate the $|V_{ub}|$ CKM matrix element and study the non-leptonic
two-body decays of beauty mesons. The authors of Ref. \cite{BLMS}
performed also a detailed study of the CP violating asymmetry for the
Cabibbo allowed two-body decays of B mesons taking into account
annihilation terms and non-factorizable contributions. 

For the hadronic matrix elements we adopt the parameterization chosen in
Ref. \cite{BSW}. In particular, for the transition between two
pseudoscalar mesons, i.e. $P_1(p)\rightarrow P_2(p^{\prime})$, one has: 
\begin{equation}
\langle P_2(p^{\prime }) \mid V^\mu \mid P_1(p)\rangle = 
(p+p^{\prime })^\mu F_1(q^2)+
\frac{m_1^2-m_2^2}{q^2}q^\mu \left[ F_0(q^2) - F_1(q^2) \right ]
\label{e:PVP}
\end{equation}
and for the transition between a pseudoscalar and a vector meson,
$P_1(p)\rightarrow V_2(\varepsilon ,p^{\prime })$:
\begin{equation}
\langle V_2(\varepsilon ,p^{\prime })\mid V_\mu \mid P_1(p)\rangle =\frac{2i
}{m_1+m_2}\varepsilon _{\mu \nu \alpha \beta }\varepsilon ^{*\nu }p^{\prime
\alpha }p^\beta V(q^2)  
\label{e:PVV}
\end{equation}
\begin{eqnarray}
\langle V_2(\varepsilon ,p^{\prime })\mid  A_\mu \mid P_1(p)\rangle
& = & (m_1+m_2)\varepsilon _\mu ^{*}A_1(q^2)-\frac{\varepsilon ^{*}\cdot q}{
m_1+m_2}(p+p^{\prime })_\mu A_2(q^2)-  \nonumber \\
&+& 2m_2\frac{\varepsilon^{*}\cdot q}{q^2}q_\mu 
\left [ A_0(q^2) -   A_3(q^2) \right] 
\label{e:PAV}
\end{eqnarray}
where $\varepsilon _\mu $ is the polarization vector of $V_2$.
The form factor $A_3(q^2)$ is given by the linear combination 
\begin{equation}
A_3(q^2)=\frac{m_1+m_2}{2m_2}A_1(q^2)-\frac{m_1-m_2}{2m_2}A_2(q^2)
\label{e:A3}
\end{equation}
and, in order to cancel the poles at $q^2=0$, the conditions 
\begin{equation}
F_1(0)=F_0(0)~~~~~~~~~~~~A_3(0)=A_0(0)
\label{e:fdfq20}
\end{equation}
should be verified.

These form factors, using the Heavy-flavour symmetry, can be related to
the corresponding form factors for $D \to K (K^*)$ transitions
\cite{IW2}. The relations were firstly obtained by Isgur and Wise in a
different parameterization for the hadronic matrix elements: 
\begin{eqnarray}
(f_{+}+f_{-})^{b\rightarrow q} &=& C_{bc} 
\sqrt{\frac{m_D}{m_B}}\;(f_{+}+f_{-})^{c\rightarrow q}  
\nonumber \\
(f_{+}-f_{-})^{b\rightarrow q} &=& C_{bc} 
\sqrt{\frac{m_B}{m_D}}\;(f_{+}-f_{-})^{c\rightarrow q}
\nonumber \\
(a_{+}+a_{-})^{b\rightarrow q} &=& C_{bc} 
\sqrt{\left(\frac{m_D}{m_B}\right)^3}\;(a_{+}+a_{-})^{c\rightarrow q}  
\nonumber \\
(a_{+}-a_{-})^{b\rightarrow q} &=& C_{bc}
\sqrt{\frac{m_D}{m_B}}\;(a_{+}-a_{-})^{c\rightarrow q}  
\nonumber \\
f^{b\rightarrow q} &=& C_{bc} 
\sqrt{\frac{m_B}{m_D}}\; f^{c\rightarrow q}\;,
\label{e:HL} 
\end{eqnarray}
where 
$\displaystyle C_{bc}=\left(\frac{\alpha_s(m_b)}
{\alpha_s(m_c)}\right)^{-6/25}\simeq 1.12$ and $q=u,d,s.$
These form factors are related to the ones defined in 
Eqs.~(\ref{e:PVP})-(\ref{e:PAV}) by the following expressions:
\begin{eqnarray}
f_{+} &=& F_1     \nonumber \\
f_{-} &=& \frac{m_1^2-m_2^2}{q^2}\left (F_0-F_1 \right )  \nonumber \\
a_{+} &=& -\frac{A_2}{m_1+m_2}  \nonumber \\
a_{-} &=& -\frac{1}{q^2}\left [(m_1+m_2)A_1^{~}-(m_1-m_2)A_2^{~}-
2m_2A_0^{~} \right ]  
\nonumber \\
f &=&(m_1+m_2)A_1 . 
\label{e:felfsta}
\end{eqnarray}

The reliability of the relations in Eq.~(\ref{e:HL}) is extensively
discussed in the literature. Quark model calculations (cfr
\cite{IsgurHL}) are consistent with them in all the semileptonic allowed
kinematical range. Corrections of order of 15\% are found for $A_1$ at
zero recoil point \cite{Dib}. Similar conclusions have been obtained by
Neubert et al. \cite{NeubertBpi} for the form factor $F_1$. Certainly
the relations in Eq.~(\ref{e:HL}) are more reliable in the zero-recoil
point, where $1/m_Q$ corrections are estimated to be small. We take
advantage of this observation to get the normalization of all $B \to L$
form factors at $q^2_{max}$ from the more recent fit of $D\to K (K^*)$
semileptonic transitions \cite{PDB}. For the sake of completeness, we
report in Tab. \ref{t:DK} the values at $q^2=0$ and at $q^2_{max}$ of
the experimentally determined values of the $D\to K (K^*)$ form factors.
In the same table the $B\to K (K^*)$ form factors at zero-recoil point
($q^2_{max}$) are also reported. 

\begin{figure}[ht]
\begin{center}
\epsfig{file=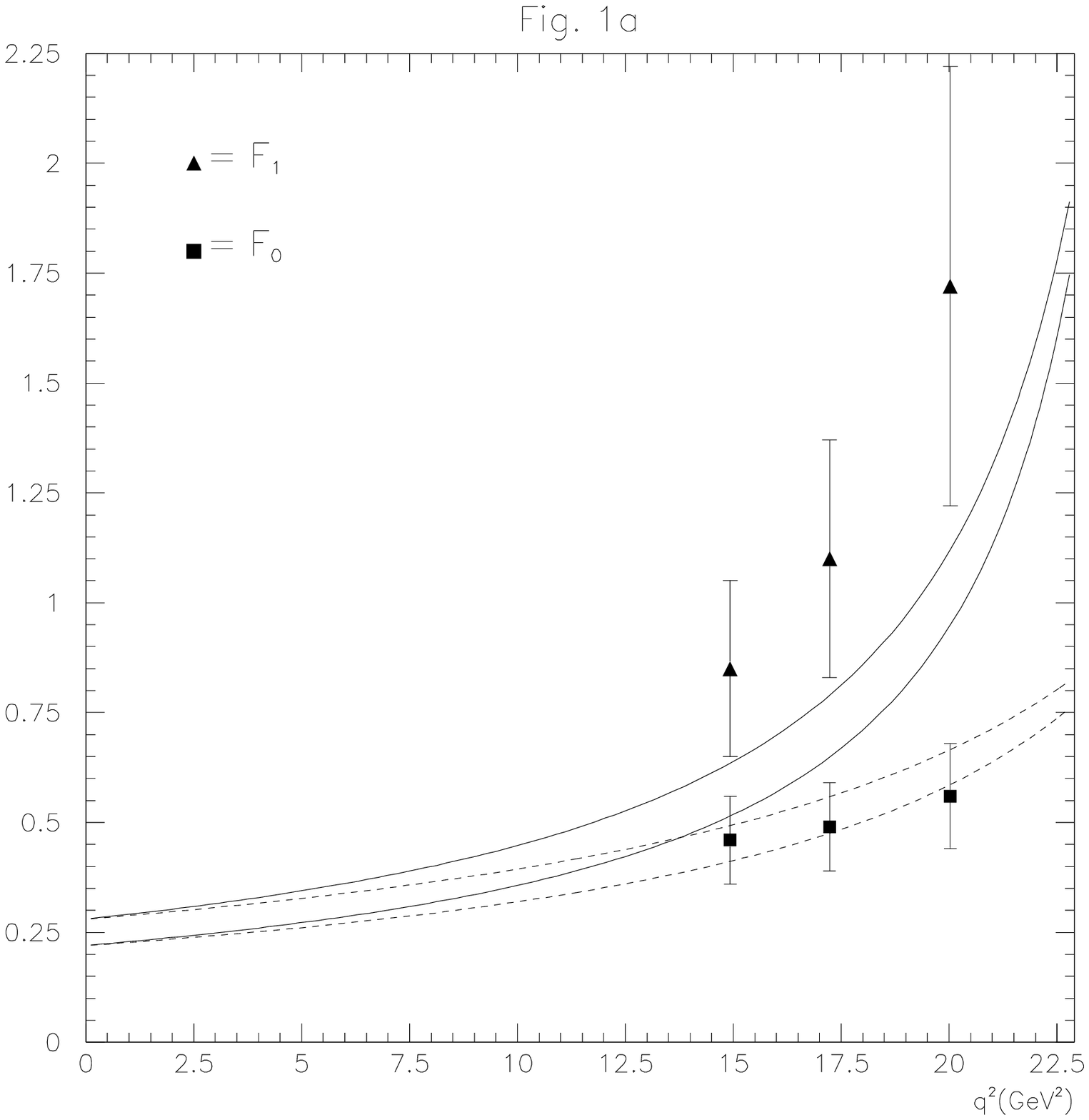,height=7.5truecm}\quad
\epsfig{file=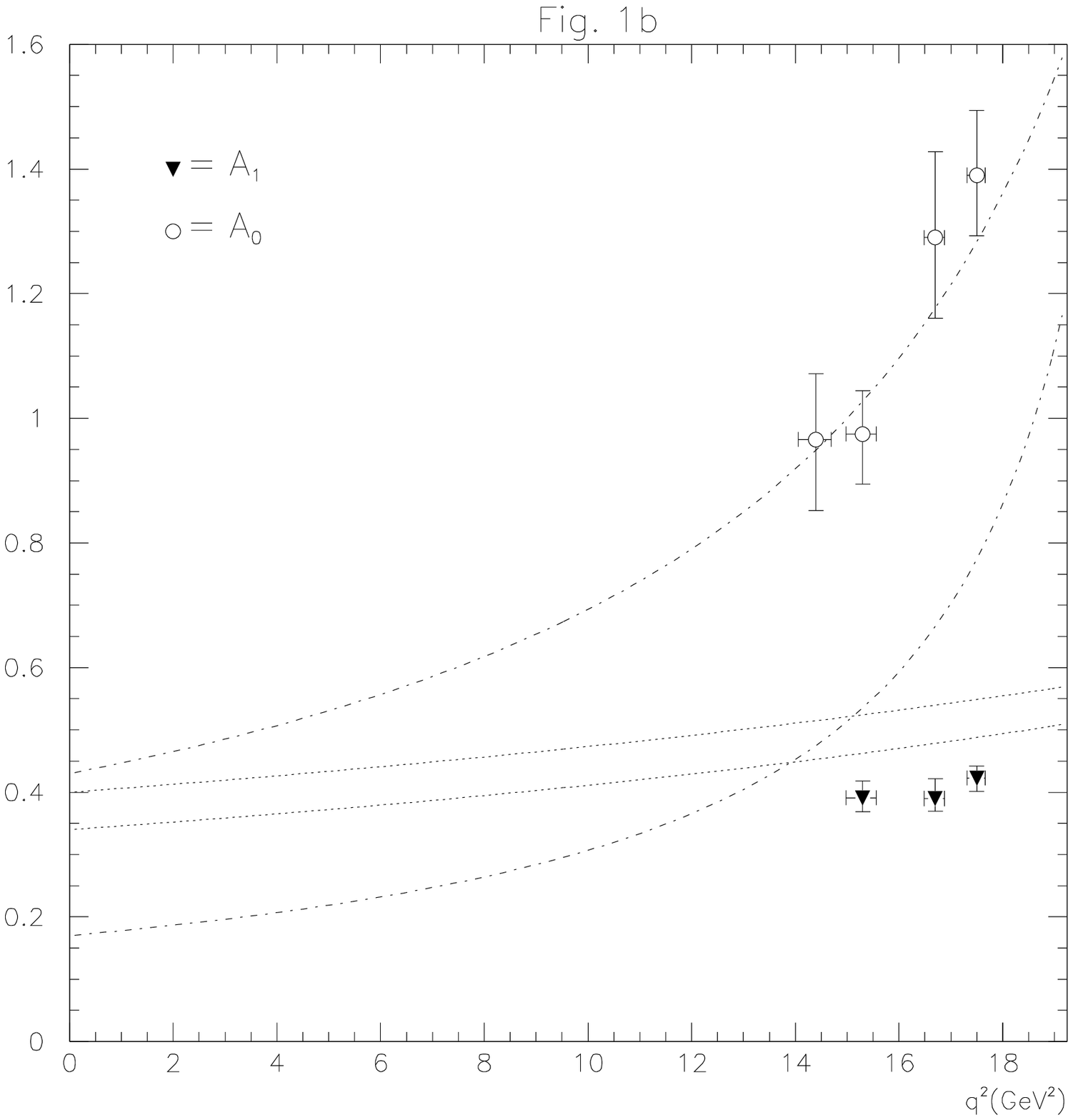,height=7.5truecm}
\end{center}
\caption{{\small 
The allowed ranges for $F^{(b\to q)}_1(q^2)$ (solid line), 
$F^{(b\to q)}_0(q^2)$(dashed line), $A^{(b\to q)}_1(q^2)$(dotted line)
and $A^{(b\to q)}_0(q^2)$(dashed-dotted line) are plotted together with
UKQCD numerical results.
}} 
\label{f:fdf}
\end{figure}

\begin{table}[ht]
\begin{center}
\begin{tabular}{|c|c|c|c||c|c|c|c|}
\hline
$~~~$ & $q^2=0$ & $q^2=q^{2}_{max,D}$ & 
\begin{tabular}{c} $M^{(cs)}_R$ \\ (GeV) \end{tabular} &
$~~$  & $q^2=q^{2}_{max,B}$ & $q^2=0$ &
\begin{tabular}{c} $M^{(bs)}_R$ \\ (GeV) \end{tabular}  \\ \hline
$F_1^{(cs)}$ & $0.74~\pm~0.03$ & $1.28~\pm~0.05$  & 2.11 &
$F_1^{(bs)}$ & $1.89~\pm~0.08$ & $0.25~\pm~0.03$  & $5.14~\pm~0.05$ \\
$A_1^{(cs)}$ & $0.55~\pm~0.03$ & $0.64~\pm~0.04$  & 2.54 &
$A_1^{(bs)}$ & $0.54~\pm~0.03$ & $0.37~\pm~0.03$  & $7.82~\pm~0.91$ \\
$A_2^{(cs)}$ & $0.40~\pm~0.08$ & $0.47~\pm~0.09$  & 2.54 & 
$A_2^{(bs)}$ & $0.71~\pm~0.08$ & $0.40~\pm~0.03$  & $6.64~\pm~0.63$ \\ 
\hline\hline
$F_0^{(cs)}$ & $0.74~\pm~0.03$ & $1.03~\pm~0.04$  & 2.57 &
$F_0^{(bs)}$ & $0.80~\pm~0.03$ & $0.25~\pm~0.03$  & $5.77~\pm~0.16$ \\
$A_0^{(cs)}$ & $0.63~\pm~0.06$ & $0.84~\pm~0.08$  & 1.97 &
$A_0^{(bs)}$ & $1.4~\pm~0.2$   & $0.30~\pm~0.13$  & $4.95~\pm~0.34$ \\ 
\hline 
\end{tabular}
\end{center}
\caption{{\small
Experimental semileptonic $D \to K(K^*)$ form factors \protect\cite{PDB}
and the values for the used resonances. The values of the $B \to K(K^*)$
form factors as results in Ref. \protect\cite{noi} are reported. 
In the last column the values, with their errors, of the resonances
compatible with the allowed ranges (see text). Note
that $A_0(0)$ is obtained by the Eq.~(\protect\ref{e:fdfq20}). }}
\label{t:DK}
\end{table}
At low $q^2$ one has the kinematical constraints (at $q^2=0$) in
Eq.~(\ref{e:fdfq20}) and the QCD sum rule calculations
\cite{Braun,Aliev,noi}. We fix the $B \to L$ form factors at $q^2=0$ in
agreement with the results of the three-points QCD sum rule approach.
These values are consistent with the corresponding results obtained with
Light-cone sum rule approach \cite{Aliev}, while different $q^2$
behaviours are predicted \footnote{Very recently a new determination of
$B \to K$ form factors was obtained in the framework of light-cone sum
rules \cite{Ball2}. The inclusion of $SU(3)$ symmetry breaking terms
allows to obtain vector form factors for $B \to K$ different from $B \to
\pi$.}. Therefore, extra assumptions are needed to cover the full $q^2$
range and we shall compare them with the results of lattice calculations
of the same form factors. Lattice calculations, in fact, provide values
for the form factors over a limited region near $q^2_{max}$. For B
decays in light mesons we assume a pole dominated model for the $q^2$
behaviour of all form factors. In Fig.~\ref{f:fdf} we plot the allowed
range of values for them compared with the UKQCD results for
$B\to\pi(\rho)$ transitions \cite{UKQCD}. 
The agreement with the lattice results is good for the vector form
factors $F_1$ and $F_0$ \footnote{Note that, using the results in Ref.
\cite{Ball2}, the agreement with the UKQCD for $F_1$ becomes better,
while for $F_0$ get worse.} and quite good for $A_0$. Large deviations
appear in the axial $A_1$ form factor: the lattice results seem suggest
large deviations from the predictions of HQS, in contrast with the
calculations of quark models \cite{Dib}. 

Looking at the Tab. \ref{t:DK} we observe that the masses of the
resonances for the $B=S=\pm 1$ mesons with the proper $J^P$ quantum
numbers agree with the mass of the lowest state in the case of negative
parity $Q\ol q$ mesons ($0^-$ and $1^-$) as suggested by Vector Meson
Dominance hypothesis. Instead, they are larger for the positive parity
ones ($0^+$ and $1^+$). 

\begin{table}[ht]
\begin{center}
\begin{tabular}{|l|c|c|c|c|}
\hline\ru1
     & \begin{tabular}{c} Experimental data \\ (in \%)\end{tabular}
& $\xi_1(w)$ &  $\xi_2(w)$ &  $\xi_3(w)$  \\ \hline \ru1
$~~~~~~~~~~a_1^{eff}$  &     & $0.963$  & $0.972$  & $0.963$  \\ 
\hline\ru1
$~~~~~~~~~~a_2^{eff}$  &     & $0.476$  & $0.443$  & $0.441$   \\ 
\hline\ru1
$Br(B^-\to D^0 \pi^{-})      $ & $ 0.53~\pm~0.05       $ & $0.497$  &
$0.509$ & $0.515$  \\ \hline \ru1
$Br(B^-\to D^0 \rho^{-})     $ & $ 1.34~\pm~0.18       $ & $1.28 $  & 
$1.28$  & $1.29$    \\ \hline \ru1
$Br(B^-\to D^{*0} \pi^-)     $ & $ 0.52~\pm~0.08       $ & $0.589$  &
$0.562$ & $0.551$    \\ \hline \ru1
$Br(B^-\to K^- J/\Psi)       $ & $ 0.101~\pm~0.014     $ & $0.0943$ &
$0.0943$ & $0.0943$    \\ \hline \ru1
$Br(\bB^0\to D^+ \pi^-)      $ & $ 0.30~\pm~0.04       $ & $0.260$  &
$0.276$ & $0.282$    \\ \hline \ru1
$Br(\bB^0\to D^+ \rho^-)     $ & $ 0.78~\pm~0.14       $ & $0.670$  &
$0.684$ & $0.690$   \\ \hline \ru1
$Br(\bB^0\to D^0 \pi^0)      $ & $< 4.8~\times~10^{-2} $ & $0.0169$ &
$0.0156$ & $0.0151$    \\ \hline \ru1
$Br(\bB^0\to D^{*0} \pi^0)   $ & $< 9.7~\times~10^{-2} $ & $0.0177$ &
$0.0163$ & $0.0158$   \\ \hline \ru1
$Br(\bB^0\to D^{*+} \pi^-)   $ & $ 0.26~\pm~0.04       $ & $0.320$  &
$0.309$ & $0.303$   \\ \hline \ru1
$Br(\bB^0\to D^0 \rho^0)     $ & $< 5.5~\times~10^{-2} $ & $0.0435$ &
$0.0412$ & $0.0403$    \\ \hline \ru1
$Br(\bB^0\to \bar K^0 J/\Psi)$ & $ 0.075~\pm~0.021     $ & $0.0907$ &
$0.0907$ &$0.0907$   \\ \hline 
\end{tabular}
\end{center}
\caption{{\small
Results of the fit of $a_1^{eff}$ and $a_2^{eff}$ to non-leptonic
Cabibbo allowed two-body decays of B mesons \protect\cite{PDB}
for different choices of Isgur-Wise function. The resulting branching
fractions are also reported.}}
\label{t:res}
\end{table}

Another test of the compatibility of our predicted range for the form
factors can be obtained by studying the exclusive non-leptonic decays.
To this extend we consider the Cabibbo allowed non-leptonic two-body decays
of B mesons. For the evaluation of the theoretical rates we adopt the
following procedure: 
\begin{enumerate}
\item
use the factorization approximation in the evaluation of the amplitudes;
\item
neglect final state interactions
\footnote{
Cfr the discussion in Ref. \protect\cite{NeuXu} for the 
validity of this approximation in the two-body non-leptonic 
Cabibbo allowed decays.};
\item
choose the following expressions for the Isgur-Wise function 
\cite{Cleo}
\begin{eqnarray}
\xi_1(w) & = & 1 - \rho_1 (w-1) \nonumber \\
\xi_2(w) & = & exp\left\{-\rho_2(w-1)\right\} \nonumber \\
\xi_3(w) & = & \frac{2}{w+1}exp\left\{(1-2\rho_3)\frac{w-1}{w+1}\right\} 
\end{eqnarray}
with
\begin{eqnarray}
\rho_1 & = & 0.91~\pm~0.21  \nonumber \\
\rho_2 & = & 1.27~\pm~0.41  \nonumber \\
\rho_3 & = & 1.53~\pm~0.50 \;;
\end{eqnarray}
\item
$a_1^{eff}$ and $a_2^{eff}$ are considered as free parameters and fitted by the 
experimental data;
\item
we fit the experimental data on these decays allowing for a variation of
the involved form factors in the previously determined ranges. An SU(3)
symmetry breaking effect in $F_1$ was considered.
\end{enumerate}

It is worth to noticing that the form factors appearing in the amplitudes 
correspond, with the only exception of $B \to K J/\Psi$ one, to the 
$b\to (u,d)$ transitions. However, we allow for $SU(3)$ breaking effects
but assume that they are small enough to be included in the obtained 
ranges. 

The compatibility of the ranges obtained before with the values of the
form factors appearing in the amplitudes is good. In Tab. \ref{t:res}
we report, for the different choices of the Isgur-Wise function, the
values of the parameter $a_1^{eff}$ and $a_2^{eff}$ and the resulting
branching ratios for the considered decays. These phenomenological
parameters contain, as discussed in Ref. \cite{Cheng}, the contribution
of the non-factorizable terms. In our approach, $a_1^{eff}$ results to
be very near to the value predicted by the analysis in
Ref.~\cite{NeubertStech} ($\sim 1$); $a_2^{eff}$, instead, results
larger than the value predicted by the same analysis ($\sim 0.3$). This
means that, in our model for the form factors, larger non-factorizable
contributions in class II and III are found. In particular our larger
values of $a_2^{eff}$ depend on the fact that the form factor
$F_1(m_{J/\Psi}^2)$ is smaller than the one in the model in Ref.
\cite{NeubertStech} so, to account for the experimental value on the
$B\to K~J/\Psi$ decay, large non factorizable contributions are needed. 

In conclusion our determination of the B form factors, based on the
relationship to the experimentally measured D form factors at
$q^2_{max}$, the QCD sum rules at $q^2=0$ and the pole dependence, are
consistent with the non-leptonic exclusive ratios and, with the
exception of $A_1$, with the lattice calculations. However, rather large
non-factorizable contributions to $a_2^{eff}$ are required. The masses 
found for the $B=S=\pm 1$ mesons, which dominate the different form 
factors, agree with the masses of the lowest mesons ($Q\ol q$ states) 
for the $l=0$, while they are larger for the $l=1$ states.

\vspace{1truecm}

We thank F. Buccella for useful discussions. One of us (P.S.) thanks
L. Del Debbio for providing us the numerical results of UKQCD 
Collaboration.

\end{document}